# Exploring Computer Science Concepts with a Ready-made Computer Game Framework

Joseph Distasio and Thomas P. Way
Applied Computing Technology Laboratory
Department of Computing Sciences
Villanova University
Villanova, PA  19085

joseph.distasio@villanova.edu
thomas.way@villanova.edu

## ABSTRACT
Leveraging the prevailing interest in computer games among college students, both for entertainment and as a possible career path, is a major reason for the increasing prevalence of computer game design courses in computer science curricula.  Because implementing a computer game requires strong programming skills, game design courses are most often restricted to more advanced computer science students.  This paper reports on a ready-made game design and experimentation framework, implemented in Java, that makes game programming more widely accessible.  This framework, called Labyrinth, enables students at all programming skill levels to participate in computer game design.  We describe the architecture of the framework, and discuss programming projects suitable for a wide variety of computer science courses, from capstone to non-major.

## Keywords
Computer science education, game programming, artificial intelligence, computer graphics, user interface design.

## 1. INTRODUCTION
There is widespread and undeniable interest in computer games on the part of computer science students, and among college students in general [7].  Computer gaming is an enormously successful industry, responsible for pushing innovation in computing and providing careers for many computer science graduates.  Responding to the demand, the inclusion of computer game programming courses has become more commonplace internationally in college curricula [2,5,6], and evidence suggests that such courses boost student enrollment and retention in computer science programs [8].

When computer game programming is taught, it is frequently in upper level or capstone courses [6,8].  The reason for delaying the use of computer gaming courses is because it requires some programming and conceptual sophistication on the part of students.  Successful game programming requires familiarity with user interface design, data structures, object-oriented design, algorithms, software engineering, graphics, artificial intelligence, and plausibly just about any other topic common to a computer science curriculum.  Thus, it is no accident that computer game courses more frequently appear later in a student's studies.  Even highly motivated students who attempt to download open-source or commercial game development frameworks from any one of many online resources [3], quickly realize that there is a steep learning curve and development of an entire game is a significant and time-consuming undertaking.  The very popular Gamelet toolkit [4] is such a framework that is ideal for experienced programmers, but is overwhelming for inexperienced programmers.  The interactive 3-D graphics software called ALICE [1] provides a fun and engaging first experience in programming for students at all levels, but using it for game programming requires significant time and experience.

The goal of the research reported in this paper is the development of a flexible, easy-to-use and compelling computer game development framework for use by all levels of computer science students, both majors and non-majors.  The Labyrinth Game Design and Experimentation Platform is a framework implemented in Java that enables instructors to expose students to very specific aspects of computer game design within the context of topics covered in specific computer science courses.  The game takes place in an underground world where a torch-carrying hero character is chased by a fire-breathing monster through a maze of tunnels.  The hero attempts to navigate the maze, lit only by the limited light from the torch, and escape before being devoured by the growling monster.  The framework enables a student to apply the ideas from computer science classes to customize this arcade-style game, which is in the genre of Pac-Man, using a level of computer programming suited to his or her ability.

This work is an extension of a student programming project originally developed as a part of a large-team "company" approach to a Software Engineering course [10].  The idea was explored further in a subsequent capstone Senior Projects course, and developed independently beyond that into its current framework as part of collaborative student-faculty research at the Applied Computing Technology Laboratory (ACT Lab) [11] at Villanova University.  That such a project can evolve from a class project to a distributable educational programming package, and can engage the attention and excitement of both student and faculty alike for two years, have been pleasant and welcome discoveries.

## 2. LABYRINTH ARCHITECTURE
The architecture of Labyrinth is modular to enable easy customization of individual components or features of the game without extensive additional programming.  Following good object-oriented design practices and software engineering conventions [9], each module handles a distinct aspect of the game's functionality.  This modularization isolates code specific to each game component in a clear and well-documented Java

API. When a student programmer wishes to explore a new idea in a certain aspect of the game, modifications are limited to the specific module that implements that functionality. Figure 1 illustrates the modular architecture of Labyrinth.

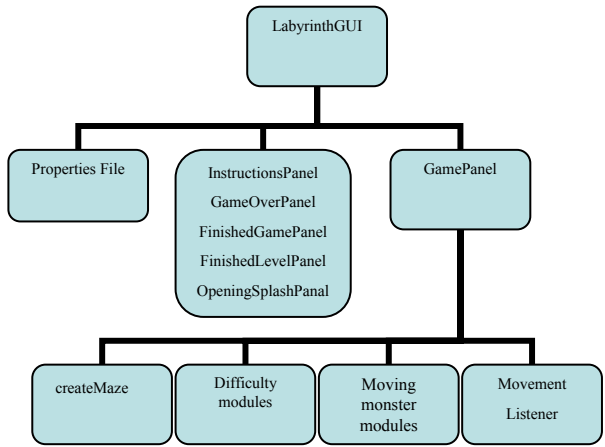

**Figure 1. Labyrinth architecture.**

The Labyrinth framework, including all source code and programmer's guide, is available for download from the ACT Lab web site [11]. The programmer's guide explains the architecture and programmer interface in greater detail than is possible here. Instructions are provided for rewriting modules, replacing modules, compiling, running, and changing the game's graphical elements, with detailed and easy-to-follow examples to assist instructors and students. The guide also points out some online resources that could be of some help to those learning how to program their own game, or seeking to make more significant modifications to Labyrinth.

## 2.1 Modules

The Labyrinth framework consists of a number of individual and disjoint modules, each of which isolate a specific feature or function of the game. These modules are:

**LabyrinthGUI module**

The first module is the LabyrinthGUI. This module is responsible for settings related to the look and feel of the game, including the Graphical User Interface (GUI) and menus. It defines the general appearance of the overall game screen. This module would be of particular interest to a project or course in graphic design or user interface design, where the students learn to use code to create images on the screen.

**Screen modules**

The next modules deal with the layout of specific screens. These modules are the InstructionsPanel, GameOverPanel, FinishedGamePanel, FinishedLevelPanel, and OpeningSplashPanel. Each of these modules is responsible for the layout of their respective screens, and would also be of interest to graphic and user interface design students.

**PropertiesFile module**

The PropertiesFile module handles mapping external resources, such as pictures and sounds, to the internal code. This module allows programmers to easily replace one resource with another, such as replacing the image of the monster with an image of a dinosaur, or the sound of echoing footsteps with footsteps in puddles of water. Projects involving this module would be of use in an introductory programming course to demonstrate the use property files and the importance of using variables for easy modification and readability.

**GamePanel module**

The GamePanel module is comprised of a collection of smaller, related sub-modules. This module controls the overall game play of the game, including all of the specific features that are handled by the CreateMaze and various Difficulty sub-modules described below.

**CreateMaze sub-module**

The CreateMaze sub-module dynamically generates the maze through which the game is played. Each maze is generated randomly, and with a very low probability that a player will ever be presented with a maze they have seen before. This sub-module would be of particular interest to an algorithm design class, offering students the opportunity to create an algorithm and implement it inside a working game. Additionally, the implemented CreateMaze algorithm is based on a common depth first search algorithm, providing students with a good starting point.

**Difficulty sub-modules**

The group of Difficulty sub-modules handles changing the level of challenge associated with game play. Difficulty is implemented as an area of visibility, a circular area of the maze that is illuminated by the hero's torch (Figure 2). There are four sub-modules in this group, one for each of the four difficulty settings: Super Easy, Easy, Medium and Difficult. Projects related to this module would be applicable in an introductory programming course, since the impact the game play is noticeable with very small modifications to the code.

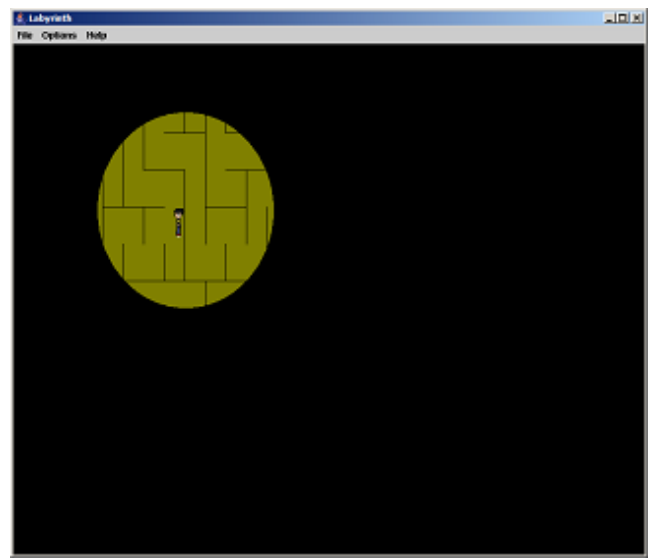

**Figure 2. Game play at medium difficulty level.**

**Monster movement modules**

The monster movement modules deal with various aspects of the behavior of the monster. This set of modules is in charge of choosing a direction for the monster, choosing an image for the monster, moving the monster, checking if the character can hear the monster, and checking if the character has been caught by the monster. Each of these tasks is implemented in its own module, with the entire set collaborating to control all aspects of the monster's movement. The logic implemented in these modules is the principle "brain" behind the game, enabling the monster to chase the hero character through the maze (Figure 3). These modules would be suited to projects in an artificial intelligence class, providing opportunities to implement a variety of concepts from game theory.

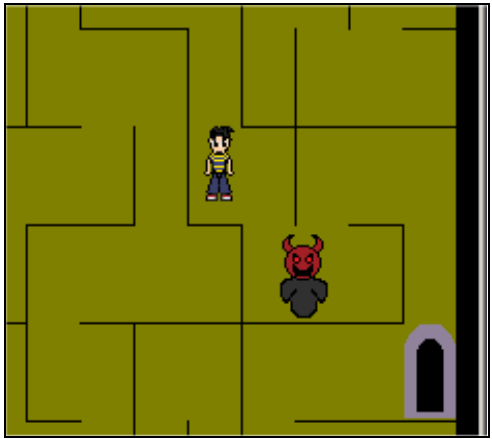

**Figure 3. Game play showing approach of monster. (close up)**

**MovementListener module**

The MovementListener module is responsible for character movement in response to player keystrokes. It maps the arrow keys to movement and direction on the game playing screen. Projects related to this module demonstrate fundamental use of listeners in Java, which would provide suitable material for an introductory programming or user interface design course.

## 2.2 Modifying Modules

The process of implementing a customized version of one of the game modules is straightforward, and is explained extensively in the bundled programmer's guide. The first step involves a code review to gain an understanding of the functionality provided by the module. Because modules range in size from a few lines to a few hundred lines of code, there is a module appropriate to any level of learning. Once the module interface is understood, the second step is to reimplement, or modify, the module to provide the desired new capability, taking care to continue to provide the same basic functionality. Code modification can either be done directly on the original module source code (i.e., commenting out the original code), or sub classing the module, overriding methods where needed. The third and final step is to recompile and test the new module.

To demonstrate the ease with which module functionality can be modified, the following example is given from the commonly and easily overridden SetEasyDifficulty sub-module. In order to replace or extend the functionality, the original code is first examined. Here is the original code from the sub-module for setting the easiest level of game play difficulty:

```
public void setEasyDifficulty()
{
  monster_speed = monster_speed_easy;
  monster_timer.setDelay(monster_speed);
  searchlight = searchlight_easy;
}
```

If, for example, game play in the new version of the game was found to be too easy at the easiest level, and a new level of difficulty was added, the modified code would look like this:

```
public void setEasyDifficulty()
{
  //monster_speed = monster_speed_easy;
  monster_speed = monster_speed_less_easy;
  monster_timer.setDelay(monster_speed);
  //searchlight = searchlight_easy;
  searchlight = searchlight_less_easy;
}
```

Replacing the graphical images used in the game, such as the hero and monster characters, can be done quite simply as well. In order to replace the images used, the user must first create new images by using a graphics drawing program. Once the new images have been made, the user edits the text-based properties file, replacing the file paths of the old images with those of the new images. Once the properties file is resaved, the game will access the new images instead of the old ones.

Modifying the GUI of the game must be done through the code modification, which requires some familiarity with Java AWT and Swing programming. The code for the GUI is located in the LabyrinthGUI module. The student can change the overall GUI look and feel, menu layout and functionality by modifying a small amount of AWT or Swing code. Changes to the GUI of individual game play and settings screens is similarly accomplished by making code modifications in one of the Screen modules.

There are a variety of tools that work quite well for modification of modules. Among the more commonly used tools are Microsoft Paint (microsoft.com) for editing images, or any of a large number of commercial or freely available and downloadable image editing tools such as PaintShopPro (jasc.com) or GIMP (gimp.org). For source code editing, basic text editors (i.e., Notepad or Wordpad) can be used, as well as freely downloadable Java programming software such as Eclipse (eclipse.org) or JCreator (jcreator.com). Labyrinth was programmed using JCreator, and a JCreator project is distributed with the source code.

Due to the computational requirements of the game graphics and movement, it is recommended that the processor speed be at least 1GHz with 256MB of RAM, so that the game can run smoothly.

## 3. CLASSROOM APPLICATIONS

Labyrinth is designed to be used in just about any computer science course to motivate learning exercises in many topics. This section describes a variety of applications for which Labyrinth is designed, with example uses of the framework within a typical sampling of computer science courses.

### 3.1 Artificial Intelligence

The monster that roams the maze in the game is an autonomous entity, controlled solely by the game program. Thus, the monster exhibits a degree of artificial intelligence, albeit rather limited. A class could learn and experiment with artificial intelligence by replacing the modules that deal with the monster's movement. There is ample opportunity for exploring more sophisticated approaches to the monster's "intelligence," such as improving how it navigates the maze, identifies dead-ends in the maze, tracks the movement of the hero, and remembers where it has been.

### 3.2 Algorithms

The majority of the modules in Labyrinth use very basic algorithms to implement the desired functionality. Since algorithm classes usually include coverage of applicable algorithms, Labyrinth could be use as a framework for engaging, hands-on experimentation with the theoretical topics covered in class. Students also could be challenged to design and implement their own algorithmic solutions within an appropriate module. The source code that is provided can be used to demonstrate practical application of algorithms discussed in class without additional implementation. For example, the CreateMaze module that is included illustrates a depth first search algorithm while generating a new maze in a straightforward and concrete way.

### 3.3 Data Structures

The Labyrinth game makes use of many different types of data structures, including some suited specifically to game play. With the data structures component of a course, Labyrinth could be used to motivate students to explore practical uses of data structures and to illustrate the importance of careful data structure design. The framework also enables a study of the more advanced data structures required in a real-world application of the sort that may be outside the scope of a textbook that focuses on fundamental issues. Practical experience in designing and implementing data structures can be motivated by the fun and immediate feedback provided by game programming.

### 3.4 Software Engineering

Because the process of software development can require more time and effort than are available in a single semester, software engineering courses wrestle with providing enough active-learning opportunities to appreciate the many process-oriented concepts covered in class. The Labyrinth framework could be used to enable focused software engineering exercise, enabling students to design and implement specific modules or extensions to the existing framework rather than undertaking the complete development of an entire game system. This targeted approach to software engineering can expose students to common, real-world challenges such as supporting and modifying OPC (Other People's Code), designing new functionality that must mesh errorlessly with an existing system, and collaborating with others in making a variety of cooperating modifications to such an existing system.

### 3.5 Theory of Computation

Labyrinth is suitable for use within a theory of computation course by providing concrete examples of the theoretical and logical concepts learned in class. Game playing theory tends to incorporate a significant amount of logic and theory in its design and implementation, and Labyrinth contains many opportunities to examine and experiment with these concepts in an engaging and hands-on way.

### 3.6 Databases

For students being exposed to the concept of databases for the first time, Labyrinth provides a simple introduction through the use of its PropertiesFile module. The flat text file used by the module and the implementation of the module can be used to illustrate basic ideas of databases in an easily accessible way, and as a foundation for subsequent database topics. For more advanced students, game playing statistics and scoring history data could be incorporated into new game functionality.

### 3.7 Computer Game Design

Obviously, a course in computer game design will expose students to a variety of theories, techniques and approaches. Labyrinth can be used as an open-source base upon which a more sophisticated game could be built, or individual functionality can be examined and re-used in brand new games. Quite often the goals of a computer game design course require students to implement a "new" game, so the strength of Labyrinth may be in providing less experienced students with a concrete example of a Java-based game that could reasonably be implemented in a single semester, given a solid effort. Alternately, Labyrinth could be used to illustrate game programming techniques, to learn about both good and bad approaches, as there are certain to be examples of both in this (or any other) computer game program.

### 3.8 Java Programming

Because Labyrinth is implemented purely in Java, it provides an engaging starting point for exploration of any level of Java programming. Examination of the code can provide clear examples of how interactivity can be accomplished in a user interface, how threads can be used to enable concurrency, and a wide variety of examples drawn from the range of problems a Java programmer must solve. Labyrinth can provide a good, running start for a larger-scale project in a more advanced course, or opportunities for an introductory level Java course to get fun and immediate feedback using minimal programming.

### 3.9 Introductory Computer Science

Introductory computer science courses for the general student population (i.e., CS0 courses) provide challenges to the instructor when programming projects are desired. Because many students may have never written a computer program, and may never do so again, finding simple projects that can motivate students is difficult. Labyrinth can provide a gentle introduction to Java at an introductory level by offering the interactivity of a computer game with the immediate feedback that comes with changing a small part. The scope of Labyrinth was purposely kept reasonable so that a student with limited programming experience could still understand it. Although the framework is advanced as a whole, many of the modules are simple by design, offering small project possibilities for introductory level computer science students.

## 3.10 Interdisciplinary Courses

One of our major goals with designing Labyrinth, and with keeping the overall architecture straightforward and easily understandable, was to foster interdisciplinary education. In an interdisciplinary computer game design course, students from many majors and with a variety of backgrounds could collaborate on the design of the game. Since most college students have experience with computer games, most will have creative ideas about the design of a game. However, since most will not have the technical expertise to implement their ideas, an interdisciplinary approach must include some students who do have programming experience. Game design teams could work much as they do at commercial game design companies. For example, English and Communications majors could create plot and story-line, Art and Design students could create graphics and an overall look-and-feel, Business majors could analyze the game design and provide input on the marketability of the idea, and Computer Science students could provide the technical know-how to implement the game itself. Of course, all team members would benefit from being exposed to both the creative team approach and to the computer science needed to make the creative ideas into a concrete finished product. Although such a course could be difficult to design and manage, we believe that there is enough depth in a variety of game design aspects to provide any student with a valuable and significant degree of learning.

## 4. CONCLUSIONS & FUTURE WORK

It is possible to include elements of computer game programming in just about any computer science course. Introductory level programming can be taught by leading students through minor modifications to existing, simple and well-organized Java code. More advanced students can reimplement entire modules and incorporate new features or better algorithms. Interdisciplinary courses can foster team-based learning, drawing from each student's area of expertise to design a customized game. Projects similar to the development of Labyrinth can make excellent software engineering or capstone projects, and can lead to continued learning beyond the end of formal coursework. Labyrinth already has been used to create humorous and fun versions, although these games have so far been restricted to a maze-running, arcade-style format. Nevertheless, the framework provides a useful teaching tool for holding the interest of all levels of students.

It has always been our hope that the Labyrinth framework would enable any college student interested in game design to try their hand at this very popular computer science discipline. Future plans for this research include deployment in a variety of projects within computer science classes (subject to agreeable colleagues), creation of a "project notebook" that will be a collection of actual and proposed programming projects suitable for a variety of courses, and continued refinement of the game including the incorporation of feedback from interested instructors and students.